\documentstyle[aps]{revtex}

\begin{document}
\draft
\preprint{}
\title{Simulation of Hard Particles in a Phase-Separating Binary Mixture}
\author{Valeriy V. Ginzburg\dag, Feng Qiu\dag, Marco Paniconi\dag\ddag, Gongwen 
Peng\dag,
David Jasnow\ddag, and Anna C. Balazs\dag}
\address{\dag Department of Chemical and Petroleum Engineering, University of 
Pittsburgh,\\
Pittsburgh PA 15261}
\address{\ddag Department of Physics and Astronomy, University of Pittsburgh, \\
Pittsburgh PA 15260}
\date{\today}
\maketitle
\begin{abstract}
We simulate the motion of spherical particles in a phase-separating binary 
mixture. By combining 
cell dynamical equations with Langevin dynamics for particles, we show that the 
addition of hard particles 
significantly changes both the speed and the morphology of the phase separation. 
At the late stage of the 
spinodal decomposition process, particles significantly slow down the domain 
growth, in qualitative agreement 
with earlier experimental data. 
\end{abstract}


Phase separation plays a significant role in determining the morphology and 
properties of polymer composites, which typically involve a blend of various 
macromolecular fluids and solid ``filler" particles\cite{VanOene}.
Despite the utility of these composites, there is little
understanding of the kinetic processes (including phase separation and wetting) 
that occur in the complex mixtures. While phase separation in binary
systems has been studied extensively theoretically and experimentally 
~\cite{Gunton83,Desai88},
the influence of solid additives on the mixtures is still poorly understood.
Recent studies have shed light on the interactions between a phase-separating 
fluid and
a stationary wall~\cite{Frisch97}, sphere~\cite{Lee} or substrate\cite{Jones91,Karim98},  but much 
less is known about the kinetics of mixtures that contain
mobile particles. To address this problem, Tanaka et al. \cite{Tanaka94} 
examined
the properties of a polymeric mixture undergoing a critical quench in the
presence of small glass particles, which are preferentially wet by one of the
components. Their results revealed that even a small concentration of hard
particles significantly changes the morphology and dynamics of the phase
separation process. However, no theoretical or computational model was developed
to characterize these changes. 

In this Letter, we report the first simulations of hard mobile particles in a
phase-separating binary mixture. Unlike earlier dynamical models of ternary
systems (developed mostly for oil-water-surfactant mixtures
~\cite{Laradji,Komura97,Kawakatsu90,Kawakatsu94}), we explicitly
take into account the "excluded volume" interaction between the particles and
the background fluid. Furthermore, we can vary the particle-fluid interactions,
 allowing for a richer range of behavior than that of a surfactant. Thus,
the model presents a new means of exploring the physical properties of
complex mixtures containing colloidal particles. Here, we consider particles
that are preferentially wet by one of the two components and show that
the boundary and ''excluded volume"
conditions at the particle surfaces significantly slow down the domain growth 
and change the morphology at the late stage of the phase separation.

We consider a phase-separating symmetric binary AB mixture that is characterized 
by the scalar order parameter $\Psi$. 
The phase separation dynamics are described by the Cahn-Hilliard equation, 
\begin{equation}
\frac{\partial \Psi}{\partial t} = \Gamma \nabla^{2} \frac{\delta {\cal 
F}}{\delta \Psi} + \xi, 
\label{Cahn}
\end{equation}
where $\Gamma$ is a kinetic coefficient, $\xi$ is a conserved zero mean Gaussian 
white noise with covariance $
\langle \xi({\bf r}, t) \xi({\bf r^\prime}, t^\prime) \rangle = - G_{1} 
\nabla^{2}
 \delta 
({\bf r} - {\bf r^\prime}) \delta (t - t^\prime)$, and
$\cal F$ is a free-energy usually given by the 
Ginzburg-Landau functional,
\begin{equation}
{\cal F} = \int\! d{\bf r} \{ - \frac{r}{2}\Psi^{2} + \frac{u}{4}\Psi^{4} + 
\frac{C}{2} (\nabla \Psi)
^{2} \}.
\label{Landau}
\end{equation}
Into this system, we introduce spherical particles of radius $R_{0}$ that 
undergo Brownian motion. The particle dynamics are described by the following 
Langevin equation,
\begin{equation}
{\bf \dot R_{i}} = M {\bf f_{i}} + {\bf \eta_{i}}, 
\label{Langevin} 
\end{equation}
where M is the mobility, ${\bf f_{i}}$ is the force acting on the i-th particle due 
to all the
other particles, and ${\bf \eta}$ represents Gaussian white noise with
$\langle {\bf \eta_{i \alpha}}({\bf r}, t) {\bf \eta_{j \beta}}({\bf r^\prime}, 
t^\prime) \rangle = 
	G_{2} \delta ({\bf r} - {\bf r^\prime}) \delta (t - t^\prime) 
\delta_{ij} 
	\delta_{\alpha \beta}$.
In this study, we neglect 
interactions between particles (i.e., ${\bf f_{i}}=0$) and only take into 
account the particles' diffusive motion. We also disregard osmotic effects 
(i.e., coupling between the particle motion and the order parameter field).

The simulation is carried out in two dimensions; our lattice is $256 \times 256$ 
sites in size, with periodic boundary conditions in both the x and y directions. 
A cell dynamical 
systems (CDS) method \cite{Oono88} is used in place of a direct forward integration 
of Eq. (\ref{Cahn}) to update the value of $\Psi$ for the 
phase-separating AB mixture. Note that $\Psi$ = 1 (-1) corresponds to the 
equilibrium order parameter for the A-rich (B-rich) phase. By employing CDS 
modeling (rather than a conventional discretization of Eq. (\ref{Cahn})), we can 
significantly increase the computational speed of the simulation. 
To simulate the particle dynamics, we discretize Eq.  
(\ref{Langevin}) and only allow the particles to move between different lattice 
sites. A ``Kawasaki exchange" mechanism is used for each particle move: first, 
the order parameter values from all the cells to be occupied by a particle in its  
"new" position are moved to the "old" position; next, the boundary and excluded 
volume conditions are imposed for the order parameter at the "new" particle position. 
This mechanism ensures the conservation of the order parameter. Such dynamics may 
break down for high particle mobilities, so we considered only the 
case where the diffusion constant is rather low 
(almost all particle "jumps" are to neighboring sites).
The 
discretized equations of motion have the following form,
\begin{eqnarray}
\Psi({\bf r}, t+1) = F[\Psi({\bf r},t)] - \langle \langle F[\Psi({\bf r},t)] - 
\Psi({\bf r}, t) \rangle 
 \rangle + \xi({\bf r}, t), \nonumber\\
F[\Psi({\bf r},t)] = f(\Psi({\bf r},t)) + D ( \langle \langle \Psi({\bf r},t) 
\rangle \rangle - 
 \Psi({\bf r},t) ), \nonumber\\
f(\Psi) = A \tanh (\Psi), \nonumber\\
{\bf R_{i}}(t+1) = {\bf R_{i}}(t) + M {\bf f_{i}} + {\bf \eta_{i}}(t),
\label{CDS}
\end{eqnarray}
where $\langle\langle*\rangle\rangle$ is the isotropic spatial 
average over the nearest-neighbor and the next-nearest neighbor sites, and 
$[ \langle\langle*\rangle\rangle - * ]$ can be thought of as a discrete 
generalization of the Laplacian.

At the surface of each particle, the lattice boundary conditions 
(specified order-parameter value and zero order-parameter flux) are imposed as:
$\Psi({\bf r},t) = \Psi_{s} $, and $ \partial_n F({\bf r},t) = 0$, 
if $R_{0} < |{\bf r} - {\bf R_{i}}(t)| \leq R_{0} + {\it a}$,
where {\it a} is the lattice spacing and $\partial_n$ denotes the ``lattice" normal 
derivative. Here, we set $\Psi_{s} = 1$ so that the particles are ``coated" by 
fluid A. The $\partial_n F=0$ condition ensures zero flux of $\Psi$ into the particles 
since $F$ plays the role of a chemical potential. 

The function $F$ in Eq. (\ref{CDS}) has a local driving term $f$ and a term
arising from the interaction with other sites; the map $f(\Psi)$ controls 
the local dynamics at each site.
It is critical that $f$ has a single unstable fixed point and
two stable fixed points symmetrically located on each side of
the unstable fixed point. Its exact functional form is not important 
for studing the universal properties of the phase separation dynamics\cite{Oono88}.
Here, we select the map $f(\Psi)=A\tanh(\Psi)$, with $A<1$ above the critical
temperature, and $A>1$ below.
 
We perform simulations for systems containing 0, 25, 50, 100, 150, 300, and 400 
particles of radius 1 
(all lengths are given in units of the lattice spacing $\it a$)~\cite{Comment2}. 
Each system was averaged over 3 runs of 20,000 time 
steps each. For all systems, the 
following values of the parameters were used: $A = 1.3, D = 0.5, G_{1} = 0, 
G_{2} = 0.5$ . The initial fluctuations 
of $\Psi$ are Gaussian with a variance of 0.05. For all runs, the composition of 
the 
fluid is fixed at 50:50 (representing a critical mixture).

The characteristic length $R(t)$ of the evolving domains is plotted as a function of 
time in Fig.~\ref{figSize}a. For $R(t)$ we use the ``broken bond" formula~ 
\cite{OJK}, $R \sim {L^{d}}/{\cal A} (t)$, where $L^{d}$ is the volume of the 
system and ${\cal A}(t)$ is the total interfacial ``area". For $d = 2$ this 
becomes,
\begin{equation}
R = \frac{L^{2}}{N_{x} + N_{y}},
\label{Perimeter}
\end{equation}
where $L$ is the system size, $N_{x}$ and $N_{y}$ are the numbers of ``broken 
bonds" 
(pairs of nearest neighbors with opposite 
signs of $\Psi$ in the x and y-directions, respectively). This measure of a 
characteristic length
empirically yields the correct asymptotic behavior for both critical and 
off-critical quenches in binary mixtures.
 
The simulations reveal that the presence of particles slows down the domain 
growth in the late stage. It can be seen that for large particle numbers ($N > 
100$), $R(t)$ undergoes a change 
from a Lifshitz-Slyozov\cite{Lifshitz61} regime, with the growth exponent of 
1/3, to a new regime. 
This new, slow-growth behavior is also characterized by a new morphology, 
different from the critical or slightly-off-critical pattern of bicontinuous 
domains.\cite{MobComment} 
The change in morphology can be seen by comparing order parameter patterns and 
particle positions 
for the system $N=300$ at times t=300 ( Fig. \ref{figN300}a) and t=3000 (Fig. 
\ref{figN300}b). 
The slower domain growth and altered structure of the mixture are qualitatively 
similar 
to the observations of Tanaka et al. \cite{Tanaka94}

In the early stage of phase separation, the formation of interfaces occurs as in 
a normal critical 
mixture, as would be expected as long as the concentration of particles is 
small, i.e., the interparticle distance is 
much larger than the particle radius. During this time, the initial domain 
growth satisfies the Lifshitz-Slyozov law, with the 
prefactor $E(n)$ smoothly depending on the particle density, 

\begin{equation}
 R(t) = E(n)t^{1/3},
\label{eq:EarlyStage}
\end{equation}
where $n = N/L^{2}$ is the particle density, and $E(n) = E_{0}(1 + \alpha n)$, 
with $E_{0} \approx 0.4$ being a growth prefactor in a particle-free system.
This growth continues until the characteristic domain size becomes comparable to 
the average interparticle distance $n^{-1/2}$. At that point, the wetting phase 
(A) percolates to form a single infinite domain (see Fig. \ref{figN300}b), and 
droplets of the B-phase are trapped
inside this domain. Further coarsening of B-domains is inhibited by the 
particles acting as obstacles to the 
motion of interfaces. 

To describe the dependence of the characteristic size on both time and the 
particle density, we use the following scaling function, 

\begin{equation}
 R(t) = n^{-1/2} (1 + \alpha n) G(b t n^{\gamma/2}),
\label{eq:Stanley}
\end{equation}
where $b \approx 1$ is a "metrical factor", $G(x)$ behaves as
$G(x) \approx G_{0} x^{1/3}$ for small $x$ ($b^{1/3}G_{0} = E_{0}$), 
and $\gamma=3$ (which is required to satisfy the transition to the 
Lifshitz-Slyozov growth law for $n \rightarrow 0$). 
There is clearly a slowing of growth 
at large time, and it is reasonable to assume $G(x) \sim x^{\delta}$ with a 
small power $\delta$ 
(or even logarithmic growth) for large $x$. 
A similar scaling form as eq. (\ref{eq:Stanley})  was used by Gyure et 
al.\cite{Gyure95} to describe the dependence of
the domain growth on the number of impurities in an Ising model. We introduce 
the additional factor $1 + \alpha n$ to account for the effective off-criticality 
induced by particles. To illustrate this scaling behavior, we plot the 
characteristic size in scaled coordinates $\rho = R(t)n^{1/2} / 
(1+\alpha n)$ vs. $\tau = t n^{3/2} $. 
For $\alpha = 13.1$, it can be seen (Fig. \ref{figSize}b) that all data fit 
reasonably well onto one master curve, with the exception of the $N=400$ case, 
where additional $n$-dependence is presumably required.  

The observed slowing down of the domain growth is reminiscent of the interface 
pinning in Ising-type systems with quenched impurities~\cite{Gyure95,Glotzer94,Huse85,Srolovitz87}.
 In all those studies, impurities reduced local interfacial 
tension and thus enforced late-time pinning, with domain growth slowing down 
logarithmically, $R \propto (\ln t)^{\eta}$. A similar effect was also seen in the "hybrid" 
model of Kawakatsu et al.~\cite{Kawakatsu94} for a binary mixture with surfactant molecules.
 On the other hand, our hard particles, which do not behave as surfactants and 
prefer to be 
in the bulk (A) phase, act as obstacles to the interface motion. When the 
characteristic domain size becomes comparable to the interparticle distance, 
interface coarsening becomes hindered by these obstacles, and the slowing down 
occurs. It is likely that at the very late stage, the domain growth would stop 
completely (as indicated by the behavior of the $N=400$ curve); however, it is 
difficult to verify this hypothesis computationally.

To elucidate the importance of wetting on the slowing down, we performed a 
simulation 
with hard, mobile particles and no preferential adsorption, i.e., with the 
following 
boundary conditions 
on the surface of the particles: 
$\partial_{n}\Psi({\bf r}, t) = 0, \partial_{n}F({\bf r}, t) = 0$.
For N=300 particles and no wetting, we observed no deviation from the 
Lifshitz-Slyozov growth law within the time scale of our simulations (20,000 
timesteps) . This result indicates that the 
 slowing down is clearly enhanced by the strong wetting and not merely by the 
effects of excluded volume or particle
 mobility. Indeed, non-wetting particles neither pin nor block interfaces, and thus, 
have only minimal effect on the dynamics of the late-stage coarsening.

Although the simulations were performed for two-dimensional systems, we believe that 
the major features (the initial off-criticality and the late-stage slowing down) would be 
found in three dimensions as well. Indeed, in 3d, the particles still would represent a network 
of obstacles for the coarsening 2d-interfaces. However, verifying this prediction remains a 
major computational challenge, even though the proposed model can be easily extended to three dimensions. 

In the computations described here, we have not taken into account 
hydrodynamic interactions or the dependence of 
viscosity on the order parameter. 
We also have not considered systematically the dependence of the growth behavior 
on such factors as temperature (which 
manifests itself in the effective diffusion constant $D_{eff}$), the clustering 
of particles that could lead to  
additional domain pinning (an effect observed by Tanaka et al. \cite{Tanaka94} 
at very high particle densities), the dependence 
on the nature of the interparticle potential and other features. These phenomena 
will be the subjects of future studies. 

D.J. gratefully acknowledges financial support from the NSF, through grant 
number DMR9217935. D.J. and A.C.B. gratefully acknowledge support from the Army 
Office of Research. A.C.B. gratefully acknowledges NSF, through grant number, 
DMR9709101, and DOE, through grant number DE-FG02-90ER45438. V. V. G. thanks 
B. P. Lee and A. Karim for useful discussions. We also thank anonymous referees 
for their helpful comments.

\newpage

\begin{figure}
\caption{ Characteristic length as a function of time: a) unscaled $R$ vs. $t$, 
b) scaled 
coordinates $\rho = R n^{1/2}/(1 + \alpha n)$ vs. $\tau = t n^{3/2}$, where 
$n = N/L^{2}$  
(data averaged over 3 runs). Growth exponent $\nu \approx 1/3$ in the early time 
range 
$10^{-2} < \tau < 10^{0}$.}
\label{figSize}
\end{figure}

\begin{figure}
\caption{Order parameter patterns and particle positions for N=300 particles at 
t=300 (a) and 
t=3000(b). 
Black circles represent particles, dark grey regions are the A-phase domains, 
light grey regions are the B-phase domains, and white points are interfaces 
($\Psi \approx 0$).}
\label{figN300}
\end{figure}


\begin{references}
\bibitem{VanOene}H. Van Oene, in Polymer Blends, D.R. Paul and S. Newman, Eds., 
Academic
Press:  Orlando (1978),Vol. 1, Chapt. 7.
\bibitem{Gunton83}J. D. Gunton, M. San Miguel, and P. Sahni, in {\it Phase 
Transition and 
Critical Phenomena}, edited by C. Domb and J. L. Lebowitz (Academic, London, 
1983), Vol. 8, 
and references therein.
\bibitem{Desai88}K. R. Elder, T. M. Rogers, and R. C. Desai, Phys. Rev. B {\bf 
38}, 4725 (1988).
\bibitem{Frisch97}S. Puri and H. L. Frisch, J. Phys. Cond. Matter {\bf 9}, 2109 
(1997).
\bibitem{Lee}B. P. Lee, S. C. Glotzer, J. F. Douglas, and A. Karim (unpublished).
\bibitem{Jones91}R. A. Jones, L. J. Norton, E. J. Kramer, F. S. Bates, and P. 
Wiltzius, 
Phys. Rev. Lett. {\bf 66}, 1326 (1991).
\bibitem{Karim98}R. Xie, A. Karim, J. F. Douglas, C. C. Han, and R. A. Weiss, 
Phys. Rev. Lett., 
{\bf 81}, 1251 (1998).
\bibitem{Tanaka94}H. Tanaka, A. J. Lovinger, and D. D. Davis, Phys. Rev. Lett. 
{\bf 72}, 
 2581 (1994).
\bibitem{Laradji}M. Laradji, H. Guo, M. Grant, and M. J. Zuckermann, J. Phys. A 
{\bf 24}, L629 
 (1991); J. Phys. Condens. Matter {\bf 4}, 6715 (1992).
\bibitem{Komura97}S. Komura and H. Kodama, Phys. Rev. E {\bf 55}, 1722 (1997).
\bibitem{Kawakatsu90}T. Kawakatsu and K. Kawasaki, Physica A {\bf 167}, 690 
(1990).
\bibitem{Kawakatsu94}T. Kawakatsu, K. Kawasaki, M. Furusaka, H. Okabayashi, and 
T. Kanaya, J. Phys. Cond. Matt. {\bf 6}, 6385 (1994).
\bibitem{Oono88}Y. Oono and S. Puri, Phys. Rev. A {\bf 38}, 434 (1988); {\bf 
38}, 1542 (1988).
\bibitem{Comment2}Although the simulation can accommodate relatively large particles,
we consider particles with the smallest possible radius in order to eliminate 
extraneous length scales and thus, more effectively analyze the dynamical 
properties of the blend. 
\bibitem{OJK}T. Ohta, D. Jasnow and K. Kawasaki, Phys. Rev. Lett. {\bf49}, 1223 
(1982).
\bibitem{Lifshitz61}I. M. Lifshitz and V. V. Slyozov, J. Phys. Chem. Solids {\bf 
19}, 35 (1961).
\bibitem{MobComment}The structure and the rate of growth are strongly dependent 
on the 
particle mobility. We find that when the particles are immobile, the slowing 
down 
begins at later times, and the characteristic domain size is usually larger than 
for mobile particles (which is consistent with \cite{Srolovitz87}).
\bibitem{Gyure95}M. F. Gyure, S. T. Harrington, R. Strilka, and H. E. Stanley, 
Phys. Rev. E {\bf 52}, 
 4632 (1995).
\bibitem{Glotzer94}S. C. Glotzer, M. F. Gyure, F. Sciortino, A. Coniglio, 
 and H. E. Stanley, Phys. Rev. E {\bf 49}, 247 (1994).
\bibitem{Huse85}D. A. Huse and C. L. Henley, Phys. Rev. Lett. {\bf 54}, 2708 
(1985).
\bibitem{Srolovitz87}D. J. Srolovitz and G. N. Hassold, Phys. Rev. B {\bf 35}, 
6902 (1987).

\end{references}
\end{document}